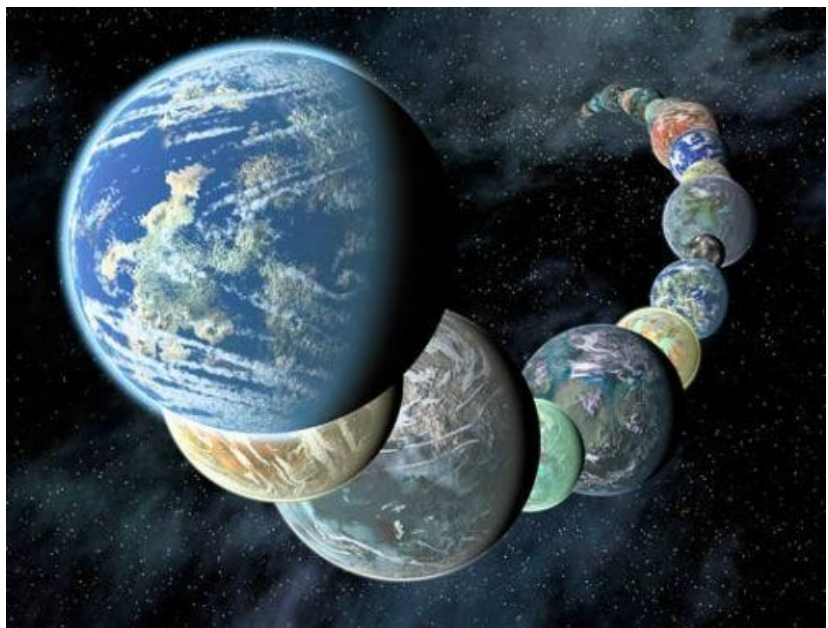

Credit: NASA/JPL - Caltech

# DIVERSITY AMONG OTHER WORLDS

# CHARACTERIZATION OF EXOPLANETS BY DIRECT DETECTION

**Schneider J.** (LUTH), Boccaletti A. (LESIA), Aylward A. (UCL), Baudoz P. (LESIA), Beuzit J.-L. (LAOG), Brown R. (STScI), Cho J. (QMUL), Dohlen K. (LAM), Ferrari M. (LAM), Galicher R. (LESIA), Grasset O. (U. Nantes), Grenfell L. (Tech. U. Berlin), Grießmeier J.-M. (ASTRON), Guyon O. (Subaru)., Hough J. (U. HertfordS.), Kasper M. (ESO), Keller Ch. (U. Utrecht), Longmore A. (ROE), Lopez B. (OCA), Martin E. (IAC), Mawet D. (JPL/ULg), Ménard F. (LAOG), Merin B. (ESAC), Palle E. (IAC), Perrin G. (LESIA), Pinfield, D. (U. HertfordS.), Sein E. (Astrium), Shore P. (U. Cranfield), Sotin Ch. (JPL/U. Nantes), Sozzetti A. (INAF-OATo), Stam D. (SRON), Surdej J. (ULg), Tamburini F. (U. Padova), Tinetti G. (UCL), Udry S. (Obs. Genève), Verinaud C. (LAOG), Walker D. (UCL/Zeeko Ltd)

## Update of a "White Paper" submitted to the ESA "ExoPlanet Roadmap Advisory Team".

### Abstract

The physical characterization of exoplanets will require to take spectra at several orbital positions. For that purpose, a direct imaging capability is necessary. Direct imaging requires an efficient stellar suppression mechanism, associated with an ultrasmooth telescope. We show that before future large space missions (interferometer, 4-8 m class coronograph, external occulter or Fresnel imager), direct imaging of giant planets and close-by super-Earth are at the cross-road of a high scientific interest and a reasonable feasibility. The scientific interest lies in the fact that super-Earths share common geophysical attributes with Earths. They already begin to be detected by radial velocity (RV) and, together with giant planets, they have a larger area than Earths, making them detectable with a 1.5-2 m class telescope in reflected light. We propose such a (space) telescope be a first step before large direct imaging missions.

## 1. OUR PHILOSOPHY: AVOID NEO-GEOCENTRISM

The thorough investigation of exoplanets will certainly constitute one of the main scientific goals for the whole $21^{st}$ century since we now know that these worlds are numerous. The challenging aims are the comprehension of the geophysics of these planets and (hopefully) the detection of biosignatures. Ultimately, this will require direct imaging with high spectral resolution and multipixel cartography. Very large interferometers will be probably unavoidable. However, in the meantime more modest instruments can still providerelevant science for giant planets and close-by super-Earths.

Exploration of Solar System bodies has revealed an extra-ordinary diversity with none object resembling to any other. The detection of the first exoplanetary systems has shown a similar diversity for mass and orbital distributions and has indicated that the Solar System is not representative of planetary systems in general. **"Diversity"** and **"Open-mindedness"** must thus be the keywords for future exoplanet exploration. Importantly, a consensus is now emerging that habitable planets are not restricted to just Earth-sized bodies (cf Workshops in AbSciCon 2008, Nantes, Aspen) : "super-Earths" with masses up to a tens of Earth's mass may well be habitable (the upper mass limit depending on local circumstances in each planetary system). Super-Earths are defined as planets for which terrestrial concepts apply: ocean/continents, planet tectonics, volcanism, habitability. It is increasingly clear that we should avoid concentrating on a new kind of **geocentrism** in which the search for "Earth's twins" is the sole ultimate goal. Stellar Physics was not triggered by the search of a "Sun's Twin".

Concerning the prospect for the number of existing planets, a recent radial velocity (RV) survey has shown that at least 30% of stars have super-Earth companions (Mayor et al 2008) and at least 10% have giant planet companions.

These starting points lead us to develop a road map in which a **single spacecraft** can provide a first "quick" characterization of both giant planets and some close-by super-Earths, by directly observing the planetary radiation (integrated over the planetary disk).

A roadmap should be inspired by science, but it finally depends on a combination of science objectives and instrumental feasibility. For now, **obtaining giant planet and super-Earth spectra is more important scientifically than obtaining the address of Earth like planets that are not yet within the possibility of direct detection**. In the following, we successively describe the scientific motivation, the instrumental concept and the feasibility of this roadmap.

## 2. SCIENCE MOTIVATION : WHAT WE WANT TO KNOW ABOUT EXOPLANETS

### 2.1 Fascinating times: a field in very rapid evolution

Presently, we have RV detection for more than 300 planets, direct detection of 8-9 bright and young giant planets (photometry only, Chauvin et al. 2005, Neuhäuser et al. 2005, Marois et al 2008, Kalas et al 2008, Lafrénière et al. 2008), radius of more than 45 transiting planets (Schneider 2008) and spectra of 2 transiting planets (HD 209458 b, HD 189733 b, (Agol et al, 2008, Deming 2008).

In a few years, CoRoT should have found several super-Earths; Kepler will confirm the proportion of Earths/super-Earths and their radius; Infrared transit surveys such as the WFCAM Transit Survey (WTS, Pinfield et al. 2008) and high-precision infrared radial velocity (e.g. NAHUAL project) will also constrain the planet fraction around M stars; HARPS S and N, EXPRESSO and CODEX will detect planets with an accuracy better than 1 m/s; SPHERE/GPI will detect young /massive (self-luminous) giants by imaging and EPICS at the E-ELT, mature (reflecting only) giants.

### 2.2 Planet characteristics

A number of planet characteristics are concerned :

*Atmosphere:* Molecules, clouds, haze, pressure and density, circulation

*Surface:* Oceans/continents, rotation, ocean glint, surface biosignatures

*Internal structure and dynamics:* For solid cores and super-Earths there are separate categories: Mercury-like (iron-rich) planet, Earth-like (rocky) planet, water-rich planet, Mini- Neptune (with primordial $H_2$ atmosphere).

*Radius & Mass:* Based on robust equation of states extended to very large pressure and temperature domains, and on various assumptions for the planetary compositions, there is a good agreement between different authors and with Uranus and Neptune observations. A power-law of the form $R/R_0 = a \cdot (M/M_0)^b$ holds with $b = 0.3$ (for a few Earth mass) down to 0.2 (up to above 70 Earth masses). The factor *a* depends strongly on the planetary composition. Information on the atmospheric composition of planets will be a key to assess the likelihood of plate tectonics. Such a process is critical for the development of life since it provides a mechanism for the recycling of volatile species on geological timescales of a few tens to a few hundreds of Myrs.

*Habitability:* It implies conditions suitable for the development of life. Since life on Earth requires liquid water, one defines the "Habitable Zone (HZ)" to be the annular region around a star in which a planet must orbit if it is to have liquid water on its surface.

Discovering water on Earths/super-Earths in the HZ will constitute a breakthrough for the general public, but scientifically will not be a surprise. Indeed, water is probably ubiquitous since it has been already detected in the atmosphere of giant planets (HD 189733 b, HD 209458 b) and in protoplanetary discs like AS 205A and DR Tau (Salyk et al. 2008).

We can thus expect that water will be detected in many planets (unless it has been photodissociated like in Venus), the exact amount depending on details of the formation scenario.

The real scientific breakthrough would be the detection of biomarkers in the HZ. These are molecules associated with life, for example nitrous oxide or molecular oxygen (and its by-product ozone). Model studies have calculated photochemical responses of these species for a range of earthlike atmospheres (e.g. Grenfell et al. 2007) but assuming a similar planetary development history as for the Earth.

**But we have no scientific grounds to estimate the probability of the presence of these molecules. It can therefore not constitute a secure goal for the first generation of instruments.**

*2.3 Derivation from observables*

The previous list of characteristics can be derived from several observables.

*Mass:* is derived from RV and astrometry, down to 1 $M_{Earth}$

*Radius:* is derived from transits for a very few planets and from the flux $F = R^2 T^4$ (thermal emission) and $F = AR^2$ (reflected light) where *A* is the planet albedo. In both cases the aid of models is necessary: adjustment of the observed spectrum to a planckian with absorption bands for the thermal flux, model for *A* for the reflected light (the fine-tuning of this model being strengthened by the polarization observations). These self-consistent models will be the result of a series of successive approximation using all observables as shown by the diagram of Figure 1.

*Atmosphere:* Planetary radiation consists of starlight reflected by the planet (at ultra-violet to near infrared wavelengths here referred to as "the visible") and of thermal emission (at infrared wavelengths). They have a different interaction with the planet (absorption and scattering, and absorption and emission, respectively). As a result, the two wavelength regions yield different, **but complementary information** about the planetary characteristics.

The IR provides ozone, $CO_2$ bands and water features. If water features can be disentangled from the wings of a Planckian spectrum, the latter give the planet temperature above clouds and the planet radius (provided the planet has no rings).

The visible provides spectral bands of water, oxygen, methane and $CO_2$ (at 1.25μ). In addition, the Rayleigh scattering gives the column density of clear atmosphere above clouds or solid surface. Another aspect of reflected light is that it is polarized. Indeed, its degree and direction of polarization depend on the planetary characteristics, on the wavelength, on the orbital phase angle (they are expected to be highest for phase angles around 90°, a favorable phase angle for direct detection). Also, polarimetry facilitates planetary detection (the star being unpolarized) and the polarized signal depends on the planetary characteristics in a different way than the flux does (Stam 2008).

The first step is to identify species thanks to spectral features in the flux and polarization. This is in general easy, unless there are blends (for instance, ozone and $CH_3Cl$ (and even silicates like diopside) have the same signature at 9.6 micron; water and $CO_2$ are blended at medium resolution at 1.25 micron). The ambiguities can be removed if the same molecule is detected at different wavelengths (e.g. water at 0.8µ and $O_3$ complemented by its parent molecule $O_2$ at 0.76 µ). Finally, haze and clouds can be identified by polarimetry in reflected light.

*Surface:* oceans and "continents" have a different temperatures and albedos and bidirectional reflection functions. They typically present a difference of ~25% in thermal flux and of a factor ~5 in reflected light. Assuming a planet with a patchy surface coverage, like the Earth, the planetary flux and polarization signals are thus modulated by the planetary rotation (Ford et al 2001, Stam 2008). The period of this rotation gives the duration of the day for these planets. Palle et al (2007) have shown that the modulation is observable in continuous photometric observation runs of a few weeks. If the rotation is known, one can then bin the spectra accordingly and search for localized biomarkers, such as vegetation signatures over continents (providing the integration time for the detection is compatible). Rotation is also important to constraint planetary formation mechanisms and to infer the possible presence of a magnetic field. Long term monitoring of the global flux can reveal icy and liquid surfaces through the glint (Williams & Gaidos 2008)

*Biosignatures:* One can search for spectral signatures of 1) organic materials themselves, and 2) by-products of photosynthetic activity. On Earth, an example of a signature of the first category is the "vegetation red edge" (VRE), a strong increase of the albedo of vegetation long wards of about 0.72 µm. Its detetectability in disk average Earth spectra has been demonstrated by Arnold et al (2002), Woolf et al (2002), and Montanes-Rodrigues et al (2006) in Earth-shine spectra. The VRE might also be detectable in polarimetry (Stam 2008). On an exoplanet, VRE-like signatures can a priori be at wavelengths different from the terrestrial 0.72 µm (Kiang et al 2007). Confusion with certain types of minerals can be avoided by comparing signals against spectral databases for minerals. Examples of signatures of the second category are spectral features of $O_2$ (at 0.76 µm), and its derivative $O_3$ (at 9.6 µm). These species are clearly visible in the Earth spectrum. Note that detection of $O_2$ and/or $O_3$ does not unambiguously indicates photosynthesis: studies have to investigate the abiotic production of these species by catalytic chemistry on solid surfaces. Clearly, the best method will ultimately be to combine both categories of biosignatures.

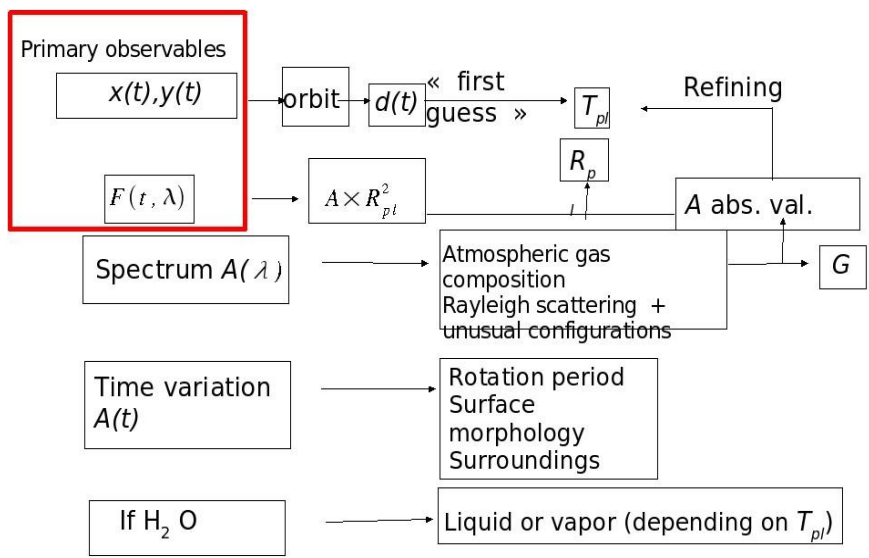

**Figure 1: Schematic self- consistent recontruction of planets properties from observables and planet models**

## 3. INSTRUMENTAL APPROACHES FOR INDIRECT DETECTION

Indirect detection are needed to provide the proportion of planets around stars, the distribution of their mass, orbit and multiplicity and for some of them targets for future target imaging. They have to be considered in the roadmap and **phased** accordingly to direct imaging missions.

*Radial velocity*

The current instrumental precision of Harps is limited to 1m/s over 4 years of data (all instrumental effects included) (Mayor et al 2008). 45 super-Earth candidates smaller than 30 Earth masses with periods shorter than 50 days have been identified. With continuous monitoring until 2017, planets with 5-10 Earth masses should be detected with HARPS in the HZ of K and G stars. The next generation spectrograph, EXPRESSO, is proposed as an instrument at the VLT. It will provide an instrumental precision of 10 cm/s for 20 minutes exposures. If selected by ESO, operations could start in 2014, allowing for the detection of planets with 2-3 Earth mass in the HZ of K and G stars. A laser-based wavelength calibration is able to reach an accuracy of 1 cm/s (Li et al. 2008), provided other sources of noise (thermal etc) are controlled at a similar level. One of them is stellar activity (acoustic modes and granulation). Acoustic noise can be reduced to 25 cm/s (observations reported in Figure 2) and in theory down to 10 cm/s in the visible range for G stars (Lovis 2007). The reduction of the granulation noise is under investigation.

Specially interesting targets are very nearby stars: a continuous monitoring of Proxima Cen has started since 5 years (Endl & Kürster 2008) and a very high cadence monitoring of alpha Cen A, B and tau Cet (1 exposure every 30 sec) has started in summer 2008 (Guedes et al 2008). In principe, a sensitivity of 1 Earth-mass is achievable in 3 years of continuous observations.

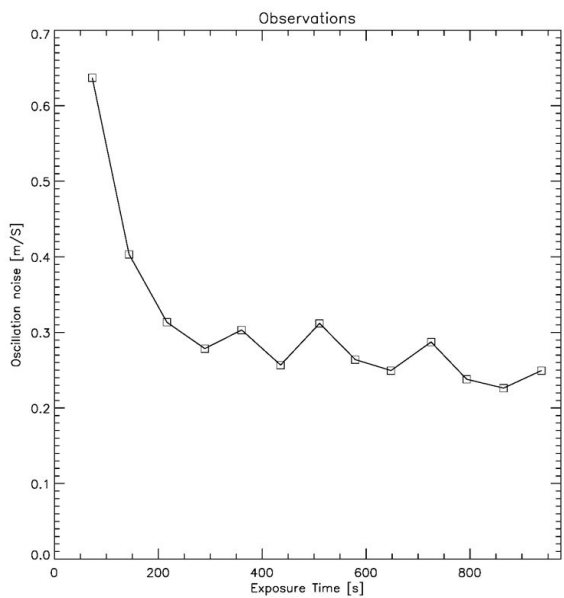

**Figure 2: Oscillation noise measured by Lovis (2007) and limiting the accuracy of RV measurements**

For M and young stars, the IR wavelength range is a better approach. It is the goal of NAHUAL (IAC www.iac.es/proyecto/nahual/), a cryogenic high resolution NIR echelle spectrograph, to be installed at the 10.4-m Gran Telescopio Canarias. It will provide 1 m/s RV accuracy for cool stars. For instance, a 10 Earth masses planet gives a RV signal of ~1 m/s at 0.3 AU of a M dwarf and is thus detectable with NAHUAL which will open a new scientific niche in the search for planets around nearby M dwarfs, young stars and red giants. A 3 Earth mass planet gives a RV signal of ~5 m/s at 0.03 UA (habitable region) of a M6 dwarf and is thus detectable with NAHUAL. It will open a new scientific niche in the search for planets around nearby M dwarfs, young stars and red giants.

An important issue is the effect of orbital inclination. For an inclination $i$ (set to 90° for edge-on orbits), the real planet mass is $M_{obs}/\sin.i$. The probability that the inclination is between 90° and $i_1$ is $P(i_1) = cos(i_1)$. For $i_1 = 30°$, corresponding to a true mass $2.M_{obs}$, the probability is $cos(30) = 0.13$. In other words, the probability that the inclination factor affects the true planet mass by a factor larger than 2 is 13%. In addition, the impact on the planet surface gravity is only a factor 1.3 (assuming a mass-radius relation $R = M^{0.3}$ and i=30°). Another way to see this is me $<M_{true}> = \sqrt{(4/3)} * M_{obs} \sim 1.15 M_{obs}$ (Griessmeier et al., 2007)..

*Transits*

The transit method can detect small sized planets, down to 1 Earth radius, especially around M dwarfs (see the Rocky Planets Around Cool Stars Network:.http://star-www.herts.ac.uk/~dpi/RoPACS). In addition, the spectroscopy of transits is a powerful "cheap" method to detect molecules in the atmosphere (Schneider 1992, Tinetti et al 2007). This method was adopted for "moderate technology" missions like CoRoT and Kepler at a time when the proportion of Earths/super-Earths was unknown. But, it has significant scientific limitations:

- the geometric probability is only ~1% for habitable planets, in other words missing 99% of habitable planets
- the transit of an habitable planet lasts 1/1000 of the orbital revolution, precluding the follow-up of evolution of planet characteristics along the orbit.
- no information is obtained on the planet surface and reflectivity in the HZ of G-K stars (for which the secondary transit is undetectable).

Transit missions limited to stars further than about 20-30 pc cannot provide targets for the first direct imaging missions. For nearer stars, TESS (NASA Phase A for a launch in 2012 - linear FOV 30 deg) and PLATO (ESA, under study for a launch in 2018) could provide a few targets for direct imaging.

*Astrometry*

Astrometry provides the inclination of each planetary orbit. Its precision is significantly less deteriorated by intrinsic stellar noise and environment (e.g., surface activity, disks) than radial velocity, thus the astrometric detection of planets in the habitable zones of solar-type stars is not severely impacted by the noise due to starspots/faculae, down to well below one Earth mass (Sozzetti 2005). While astrometry does not allow to directly obtain any further information on crucial structural properties of a planet (e.g., its radius and atmosphere), neither it can provide any clue on its atmospheric composition, it has the potential to help direct detection missions in carefully predicting times of maximum brightness, based on the full orbit determination.

The projected capabilities of future astrometric observatories such as Gaia and SIM/SIM-L aiming at micro-arcsecond precision, hold the promise to detect and characterize orbits of tens of thousands of massive exoplanets, hundreds of massive multiple-planet systems, and to be sensitive to planets as small as Earth in the Habitable Zone of the nearest 50-100 stars. A double-blind test campaign is presently undergoing to test the possibility of detecting Earth-mass planets in multiple systems in different configurations using a combination of SIM astrometry and high-precision RVs. However, astrometry may not require a dedicated mission. Indeed, a 8 m class single aperture coronagraph equipped with a 4x4 arcmin camera has an astrometric capability with a precision of 1 µas (Brown 2008).

From a strategic point of view, astrometry has two functions: detect targets for direct characterization and give the planet masses. RVs can almost do the job, at least for super-Earths.. Indeed, in spite of the fact that RV does not give the exact orbital required to disentangle the M.sini product, the probability that a planet has a true mass M twice the M.sini product is only 13%. The impact on the planet surface gravity $MR_{pl}^{-2}$ is a factor ~ 1.3.

*Micro lensing*

Microlensing is well adapted to detect planets at 0.5 – 3 UA of the parent star (projected distance on the sky) for M to F stars, and is sensitive down to 1 Earth mass. It has presently detected the smallest planet around an M star. There is a good opportunity to have planetary microlensing events as a by-product of the EUCLID mission. Microlensing cannot provide targets for future direct imaging but an estimate of the planet fraction (> 1 Earth mass, a > 1 AU).

## 4. DIRECT IMAGING

Direct imaging is definitely unavoidable to measure spectra and polarimetry of exoplanets. Combined with RV, a single image gives the orbital inclination and hence the dynamical mass. An instrument for direct imaging can also make spectroscopy of transits even for unresolved systems using its photometric capability (like with Spitzer and later JWST).

While reflected light can be detected with a coronagraph, an external occulter or a Fresnel imager, thermal IR requires an interferometer to angularly separate the planet from its parent star. In both cases a stellar light suppression mechanism is required. For a long time, the mid IR was considered more favorable for several reasons. One is that the star/planet ratio is 1000 times lower than in the visible and near IR. However, what really matters is the ratio of the planet intensity to that of the stellar residuals/leakage. Laboratory prototyping has shown that this ratio is identical in the visible and thermal IR. Interferometer prototypes are now reaching rejection rates of $10^5 - 10^6$ while coronagraphs are achieving contrasts of $10^8$-$10^{10}$. The net result is that the effective planet to stellar speckle ratio is about the same in both cases. **There is thus, from this point of view, no advantage of one approach over the other.**

*4.1 On-going direct detection projects*

On the ground, two planet finder instruments (**SPHERE, GPI**) on 8-m class telescopes will see first light in 2011. They are optimized for low resolution spectroscopy in the near IR (0.95 – 2.3 µm) and will have the capability to characterize 3 class of planets: very young giant gaseous planets at 50-150pc, intermediate age giants planets, and some mature but massive nearby planets. SPHERE has also a visible capability to detect (no spectra) in some favorable cases some Jupiter-like planets reflected light for very bright and nearby stars (<10pc).

The next generation of planet finders on the ground will be installed on **Extremely Large Telescopes** presumably in 2018-2020. The main objectives will be the study of mature giant planets (possibly down to Jupiter mass) in the near IR at low resolution and will tentatively seek for telluric planets like Super Earth. Like SPHERE, there will be visible camera but without spectroscopic capabilities.

In space, the next precursor is definitely **JWST** which embarks several coronagraphs in the near and mid-IR with the objective to acquire broadband photometry of mature giants in direct imaging. JWST may have also the ability to study transiting telluric planets as it was made with Spitzer (Tinetti et al. 2007).

The characterization of visible spectra of mature giants and massive telluric planets is not planned with current projects and remain a niche for a spatial coronagraphic telescope.

**Note that, for these facilities, telescope time has to be shared with general astrophysics, especially cosmology. In addition, from the ground it is not possible to monitor continuously a planet for several weeks.**

*4.2 Related additional science of direct imaging: Protoplanetary and debris discs*

Debris disks (like beta Pictoris) or protoplanetary discs (like LkH alfa 300) will be easily detected by direct imaging. In addition, the observation of gaps in these discs is an indirect evidence for planets. The physics of disks is relevant to understand the planetary formation stage. A coronagraphic single aperture telescope is definitely the best approach to reveal the complex structures of such extended objects. The identification of structures is essential to understand the disk/planet interactions. A number of additional science that is not listed here can benefit from a space telescope. Finally, when a gap in the disk is due to a planet, its structure provides a significant constraint on the mass of the planet. This approach has been successfully applied to the planet Fomalhaut b (Chiang et al. 2008). It is of particular interest when the planet mass cannot be derived from radial velocity or astrometric measurements. This is the case when the star is too active, too hot or rotating too fast or when the planet is too far away from the parent star to exercise a gravitational pull on it.

**5. REQUIREMENTS FOR INSTRUMENTAL IMPLEMENTATION OF A 1.5-2 M CORONAGRAPH**

*Basic concepts*

Based on astrophysical requirements the technical aspects of a mission (like the SEE-COAST proposal to Cosmic Vision) to detect and characterize exoplanets down to close-by Super-Earths can be derived. The philosophy is a simple and compact telescope and spacecraft to reduce cost and development time, the complexity and requirements for achieving high contrast being relayed to the focal instrument. Here, a 1.5-2 m class off-axis telescope is needed to reach a reasonable amount of targets (giant planets and close-by super-Earths) provided high contrast can be obtained at 2 or 3 lambda/D between 0.4 and 1.25 µm.

For the focal instrument, spectroscopy and if possible polarimetry are necessary. It is probably necessary to separate the visible and near IR parts into 2 channels for technical issues (detectors, coatings, dimensioning, etc). Here, spectroscopy refers to spectro-imaging to maximize detection capabilities in a reasonable field of view. For that, the heritage of SPHERE and EPICS will be a major advantage for the technical aspects and instrumental modelling.

In the previous SEE COAST study we identified 3 main critical aspects or sub-systems, namely: 1/ the wavefront errors (WFE), 2/ the system to suppress the starlight (coronagraph), and 3/ the system to perform calibration and hence further improvement of the contrast.

*Wavefront requirements*

The primary requirement of high contrast imaging is a good optical wavefront. The telescope mirrors (especially the primary) need to be super polished. Some developments are on-going for off-axis designs. It would be necessary to reach a level of about 5nm rms (lambda/100) at mid-frequencies. This aspect is challenging but not un-feasible. An alternative or complementary approach is to correct the wavefront. In this case a sensor and a corrector are needed. Several concepts of focal plane sensors are being developed especially in Europe in the context of EPICS or for general R&D. The implementation of a Deformable Mirror (DM) in a space telescope needs to be assessed. A study was made in the context of TPF-C. It concluded that a DM can be spatially implemented in a warm telescope. According to this study, warming the telescope has also the advantage of minimizing thermal drifts. Japanese teams are pursuing a different approach for the SPICA coronagraph and are now prototyping a cryogenic DM. Whatever the solution, it clearly demonstrates that wavefront correction for space telescope is becoming feasible and should be considered in our approach for assessment. The SPICA project would in this respect constitue an excellent precursor. On our side, we are developing a new promising concept of wavefront measurement, the "Self-Coherent Camera" (Baudoz et al. 2006, Galicher et al. 2008) illustrated in Figure 3. Another solution is the single-mode pupil remapping (Lacour et al 2006), which will be tested on the sky at the end of 2008.

In addition to phase defects, the instrument will suffer from amplitude defects and it is important to assign a specification on this to make sure it is fully assessed. For that, a rigorous analysis of the optical system is needed.

*Coronagraphy & starlight suppression*

The off-axis telescope design has been chosen to satisfy the requirements of a coronagraph since these are mostly sensitive to pupil obscuration. The performance of a coronagraph is intimately related to the wavefront quality. However, it should also provide a high throughput (>50%), large discovery space, small Inner Working Angle (IWA) of 1-2 lambda/D and be insensitive to chromatism.

Several concepts exist, many are being prototyped across the world, some have already demonstrated capabilities that are compatible with our goal. Among them, the Annular Groove Phase Mask (Mawet et al. 2005), the Multi-Stage 4 Quadrant Phase Mask (Baudoz et al. 2007, Figure 3), the Phase-Induced Amplitude Apodization Coronagraph (Guyon et al. 2005) are quite promising. Let us also mention optical vortices (Mawet et al 2005, Anzolin et al 2008) .

The contrast requirement should be compatible with the flux ratio of the planets we are looking at, ~$10^7$-$10^8$. Such performance have been met already in laboratory demonstrations (Trauger & Traub 2007, Baudoz et al. 2007, 2008).

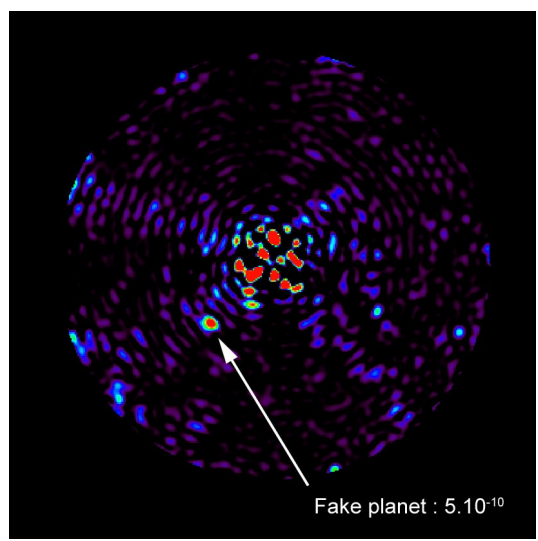
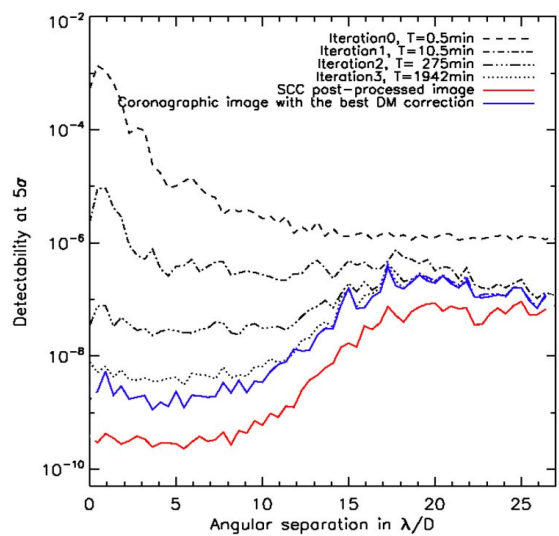

**Figure 3: Laboratory experiment of a multi-sage phase mask (left) obtained at the Observatoire de Paris and simulation of detectability with a Self Coherent Camera combined with a space 1.5 m coronagraph (right).**

*Further speckle rejections*

In addition to wavefront quality that ensures a low speckled halo, additional rejections are feasible by the use of differential imaging e.g. by using spectral, polarimetric or coherence characteristics or possibly a combination of them. These techniques can be directly inherited from the SPHERE and EPICS projects. A significant gain is to be expected although a thorough system design compatible with such techniques would be required.

*Preliminary performances*

At the time of the COSMIC VISION proposals, an intensive simulation work involving an instrumental model was carried out for SEE COAST to address contrast performances and the statistics of the detections (Figure 4). As a result it is shown that such a mission has a huge capacity to characterize giant planets and to obtain spectra for the first time in the visible. A significant fraction of close-by Super Earths spectra is also detectable.

The development of a end-to-end simulation code is now mandatory to address the tolerances of critical elements (coronagraph position for instance) and then to put specifications on the system.

*Cost estimate*

A preliminary cost was assessed for the COSMIC VISION proposal and was found to be compatible with the M class mission of ESA (300 M€) The instrument itself was only 5% of this cost while most of the cost is driven by the telescope, the platform and the launch. This clearly supports our strategy to allow for more complexity in the instrument rather than at the telescope or platform levels.

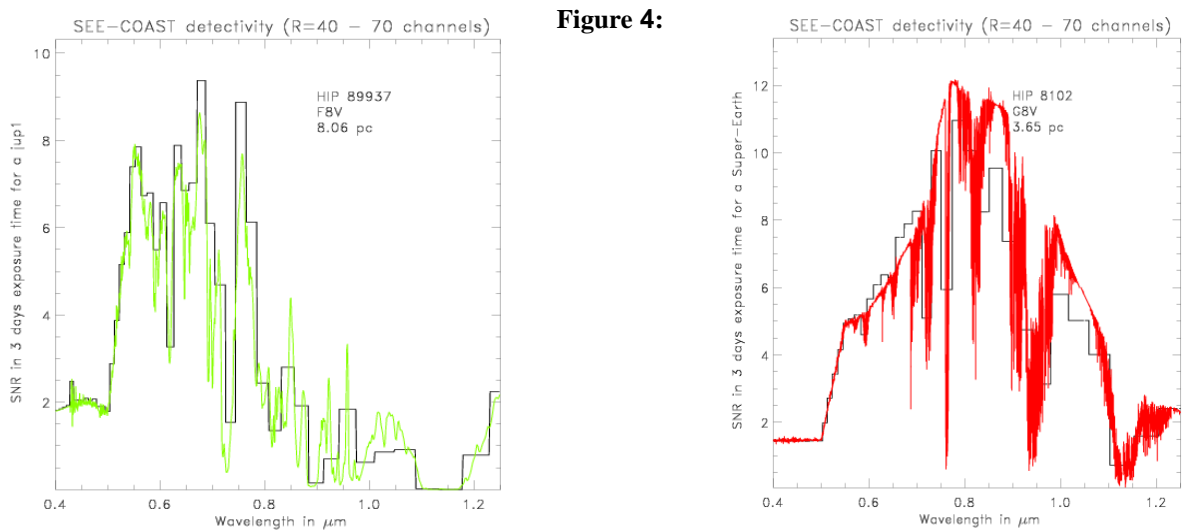

**Figure 4:**

**Instrumental simulations were made for COSMIC VISION to assess the capabilities of SEE COAST. Two examples are shown for illustration here for a mature Jovian planet (left) and a Super Earth of 2.5 Earth radii (right).**

## 6. EUROPEAN COMPETENCE

Technological developments for space missions geared to the detection of exoplanets were started in Europe by several institutes and industrials. A single aperture mission can benefit a lot from the GAIA heritage (and certainly other missions).

For Gaia, a full SiC mirror with 1.5 x 0.5 m pupil dimension is currently being developed. The optical quality of the instrument reaches a wavefront error of 50 nm. Thanks to a complex thermal control and the high thermal conductivity of SiC, thermal gradient are controlled in the range of a few milli Kelvin at 130 K during 6 hours. It is a good starting point for the development of a coronagraphic instrument. In order to reach the challenging requirements of the coronagraphic instrument, several elements are to be improved for both ground and operational phases
- enhanced WFE control during polishing and test phases
- enhanced thermal control of the payload (more efficient thermal shield)
- active optics on secondary mirror to improve the optical quality during manufacturing . In that case an efficient WFE sensor has to be implemented on the telescope

The pointing accuracy and stability (0.5 mas to 2mas) required to maintain the coronagraph on the star is also comparable to Gaia performances. Indeed, for Gaia it would be possible to measure the centroid position of a star with an accuracy of 0.2mas for m =15 stars. This accuracy includes all contributions (non uniformity of the detector response, stability of the telescope…).

High accuracy proportional thrusters are currently developed for Gaia. It has to be checked that the current cold gas thruster configuration is sufficient. FEEP thrusters technology could be an alternative solution.

Several laboratories, like the National Centre for Precision Surfaces at the University of Cranfield (UK), Zeeko Ltd (UK) and LAM (France) have the competence to high precision polishing (10 nm WFE). New stress-polishing techniques allow generating highly aspherics optics, including off axis mirrors, free of the usual mid- and high spatial frequencies errors, using a pure spherical polishing with full-size tools. In this framework, a common patent between the SESO Company and LAM has been recently registered for the manufacturing of large (1.5 – 2 m) aspherical off-axis mirrors. The feasibility of an active corrector solution has also been investigated recently by a LAM/THALES-Alenia-Space study.

## 7. CONCLUSIONS

Since a scientific progression is necessary to understand the physics of planets (first giant planets and super-Earths, then Earths) we naturally derive a series of conclusions:

- **A strong support to high cadence high accuracy RV searches is considered a priority**.

- Astrometry might be necessary to find Earth-like planets (if their frequency is small) but is not necessary for Super Earths (RV will take care of this task). Therefore, **an astrometric mission is not recognized as a first priority**

- A direct imaging mission in the visible fills the gap between current projects and future missions. **Spectro-polarimetric characterization of mature giants and Super Earths will be unique**. Earth-sized planets, being more difficult to detect, will deserve more astrophysical/instrumental developments. But we do recognized that MIR and Visible are both necessary. In that respect a MIR interferometric concept (Darwin/TPF-I) and large 4m-class single aperture missions for the visible are proposed as further steps.

- Only a space mission (contrary to ground-based instruments) can monitor a planet for several weeks as required by a thorough characterization.

Consequently, we are then naturally led to propose the roadmap given by Table 1 where steps in blue are already ongoing or approved (including ELTs).

Table 1:Successive phases of the proposed roadmap

| Projects | Spectral bands | Outputs | Planets |
|---|---|---|---|
| **1/ High Accuracy RV** | Visible/NIR | Mass, address, statistics | Giant and super-Earths |
| **2/ SPHERE/GPI (2011)** | NIR | Photometry & spectra | Young/massive nearby giants |
| | Visible | Photometry, polarization | Young/massive nearby giants |
| **3/ JWST (2013)** | NIR | Photometry & transits | Young/massive nearby giants |
| | MIR | Photometry & transits | Young/massive nearby giants |
| **4/ SPICA (2018)** | NIR-MIR | spectroscopy & transits | Young/massive nearby giants |
| **5a/ ELTs (2018-2020)** | NIR | Spectroscopy | Mature giants, super-Earths? |
| | Visible | Photometry, polarization | Mature giants, super-Earths? |
| **5b/ SEE COAST (1.5 – 2 m)** | Visible / NIR | photometry & spectra & degree of polarization | Mature giants, nearby super-Earths |
| **6/ Astrometry / RV with ELT** | Visible | Mass, address, statistics | 1 Earth sized planets |
| **7a/ DARWIN / TPF-I** | MIR | Spectra | 1 Earth sized planets |
| **7b/ TPF-C** | Visible / NIR | Spectra | 1 Earth sized planets |
| **7c/ TPF-O / Fresnel** | Visible / NIR | Spectra | 1 Earth sized planets |